\documentstyle[12pt,version2,aps]{revtex}
\begin{document}
\renewcommand{\theequation}{\arabic{section}.\arabic{equation}}
\newcommand{\eqreset}{\setcounter{equation}{0}}
\setlength{\textheight}{20cm}
\vspace*{.9 in}
\begin{center}
{\large\bf
Quantum Heisenberg Chain \\
with Long-Range Ferromagnetic Interactions \\
at Low Temperature }
\vspace{.5 in}\\
 Hiroki {\sc Nakano}
         and
    Minoru {\sc Takahashi}
 \vspace{.3 in}\\
        {\it Institute for Solid State Physics,University of Tokyo}\\
        {\it Roppongi,Minato-ku,Tokyo 106} \\
(received ~~~~~~~~~~~~~~~~~~
\end{center}

\begin{abstract}

    A modified spin-wave theory is applied
to the one-dimensional
quantum Heisenberg model with long-range ferromagnetic interactions.
\newline
\mbox{(${\cal H} = - J \sum_{i < j} (r_{i j})^{- p}
\mbox{\boldmath $S$}_{i}\/\cdot\/ \mbox{\boldmath $S$}_{j} $)}.
Low-temperature properties of this model are investigated.
The susceptibility and the specific heat are calculated;
the relation between their behaviors
and strength of the long-range interactions
is obtained.
This model includes both the Haldane-Shastry model and
the nearest-neighbor Heisenberg model;
the corresponding results in this paper are in agreement
with the solutions of both the models.
It is shown that
there exists an ordering transition for $1<p<2$
where the model has longer-range interactions
than the HS model.
The critical temperature is estimated.

KEYWORDS: Heisenberg chain, long-range interactions, Haldane-Shastry model,
modified spin-wave theory, susceptibility, specific heat, critical temperature
\end{abstract}

\section{Introduction}\label{intro}

    The quantum Heisenberg chain with long-range interactions
decaying as $1/r^{2}$, which is called the Haldane-Shastry model (HS),
has been extensively investigated
since the exact eigenstates and their eigenenergies are obtained independently
by Haldane\cite{Haldane1} and Shastry.\cite{Shastry}
The thermodynamics of this model was investigated by Haldane.\cite{Haldane2}
Unfortunately few studies for the general type of $1/r^{p}$ have been made.
What we know are
rigorous bounds for the correlation functions in the disordered phase,
which are established by Ito.\cite{Ito}

    On the other hand, the model of the classical spins with long-range
interactions has been studied for about twenty years.
In the case of 1-component spins (Ising model),
it was proved by Dyson\cite{Dyson}
that the model in the region $ 1< p < 2 $ has an ordering transition and
that the model in the region $p > 2$ doesn't have it.
In the case of 2-component spins (XY model),
\v{S}im\'{a}nek\cite{Simanek} showed that
with the low-temperature harmonic approximation
there exists a Kosterlitz-Thouless-like transition
to a low-temperature phase with infinite susceptibility.
In the ferromagnetic case of 3-component spins (Heisenberg model),
simulation of the case when $p = 2$
was done by Romano\cite{Romano} with the Monte Carlo method.
The renormalization group approach
was made by Fisher et al.\cite{Fisher,MSuzuki}

   In comparison with the model with long-range interactions,
the Heisenberg model with a short-range interaction i.e.
a nearest-neighbor interaction (NN Heisenberg model) has a longer history.
It is well known that for $S=1/2$ the exact solution is obtained
with the Bethe ansatz method.
The modified spin-wave approximation was first proposed
by one of the authors.\cite{mspin}
The conventional spin-wave theory cannot be applied
to the case when the dimensions are less than three.
This modification makes the spin-wave theory valid
even in one and two dimensions.
Results from the modified theory are in good agreement with
those from the Bethe ansatz integral equations.\cite{TY1,TY2}

  In this paper we consider the quantum Heisenberg model
with long-range ferromagnetic interactions decaying as $1/r^{p}$
in one dimension.
Its Hamiltonian with a periodic boundary condition is written as follows;
\begin{equation}
\label{hamiltonian}
{\cal H} = - \frac{1}{2}
\sum_{m=1}^{N} \sum_{n=1}^{N - 1} J
[\frac{\frac{\pi}{N}}{\sin (\frac{\pi n }{N})}]^{p}
\mbox{\boldmath $S$}_{m} \cdot \mbox{\boldmath $S$}_{m+n}
\end{equation}
This model in the limit $p \rightarrow \infty $ is the NN Heisenberg model and
this model for $p = 2 $ is the HS model.
We discuss a modified spin-wave theory of this model
and study the properties at low temperature.
In the next section a formulation of the modified spin-wave theory is done.
In \S 3 the terms up to the second order of Bose operators
in the transformed Hamiltonian are considered.
In the region $p \ge 2$
the temperature-dependence of the susceptibility and of the specific heat
are calculated;
the critical temperature in the region $1 < p < 2$ is estimated.
In \S 4 the terms up to the forth order of operators are considered.
The procedure to obtain physical quantities is shown.
The susceptibility and the specific heat are analytically obtained for $p = 2$.
In \S 5 we discuss the results.
They are compared to the solutions of the HS model and
of the NN Heisenberg model.

\eqreset
\section{Formulation of Modified Spin-Wave Theory}\label{form}

   First the Holstein-Primakoff transformation
\begin{eqnarray}
S_{m}^{+} &=& S_{m}^{x} + i S_{m}^{y} = \sqrt{2 S} f_{m}  (S) a_{m}
\nonumber \\
S_{m}^{-} &=& S_{m}^{x} - i S_{m}^{y} = \sqrt{2 S} a_{m}^{\dagger} f_{m}  (S)
\nonumber \\
S_{m}^{z} &=& S - a_{m}^{\dagger} a_{m}
\nonumber \\
f_{m} (S) &=&  \sqrt{1-(\frac{1}{2 S}) a_{m}^{\dagger} a_{m} }
= 1 - \frac{1}{4 S} a_{m}^{\dagger} a_{m} + O(S^{-2})
\end{eqnarray}
is applied
to the Hamiltonian (\ref{hamiltonian}).
Expanded with respect to $1/S$, the Hamiltonian is rewritten as follows:
\begin{eqnarray}
{\cal H} &=& E_{0} + {\cal H}_{2} + {\cal H}_{4} + O(S^{-1}),
\label{hamiltonian2} \\
E_{0} &=& - \frac{1}{2}
\sum_{m=1}^{N} \sum_{n=1}^{N - 1} J S^{2}
[\frac{\frac{\pi}{N}}{\sin (\frac{\pi n }{N})}]^{p},
\nonumber \\
{\cal H}_{2} &=& \frac{1}{2}
\sum_{m=1}^{N} \sum_{n=1}^{N - 1} J S
[\frac{\frac{\pi}{N}}{\sin (\frac{\pi n }{N})}]^{p}
(a_{m+n}^{\dagger}- a_{m}^{\dagger})(a_{m+n}- a_{m}), \nonumber \\
{\cal H}_{4} &=& \frac{1}{8}
\sum_{m=1}^{N} \sum_{n=1}^{N - 1} J
[\frac{\frac{\pi}{N}}{\sin (\frac{\pi n }{N})}]^{p}
\{ a_{m}^{\dagger} a_{m+n}^{\dagger} (a_{m}- a_{m+n})^{2} \nonumber \\
& & \ \ \ \ \ \ \ \ \ \ \ \ \ \ \ \ \
+ (a_{m}^{\dagger}- a_{m+n}^{\dagger})^{2} a_{m} a_{m+n}  \}. \nonumber
\end{eqnarray}

   Next the site representation is changed to the momentum representation
with the Fourier transformation
$a_{m}=(1/\sqrt{N}) \sum_{k} e^{i k m} a_{k} $ ,
$a_{m}^{\dagger} = (1/\sqrt{N}) \sum_{k} e^{- i k m} a_{k}^{\dagger} $.
Then ${\cal H}_{2}$ and ${\cal H}_{4}$ are transformed
to the following equations:
\begin{eqnarray}
{\cal H}_{2} &=& \sum_{k} a_{k}^{\dagger} a_{k}
S \{ \eta (0) - {\rm Re} [ \eta (k) ] \}, \
\eta (k) \equiv J \sum_{n=1}^{N - 1}
[\frac{\frac{\pi}{N}}{\sin (\frac{\pi n }{N})}]^{p} e^{i k n} \nonumber \\
{\cal H}_{4} &=& \frac{1}{8 N}
\sum_{k_{1},k_{2},k_{3}}
a_{k_{1}}^{\dagger}a_{k_{2}}^{\dagger}a_{k_{3}}a_{k_{1}+k_{2}- k_{3}}
\{ \eta (k_{1}) + \eta (- k_{2}) + \eta (- k_{3}) \nonumber \\
& & + \eta (k_{1} + k_{2} - k_{3})
- 2 \eta (k_{1} - k_{3}) - \eta (k_{3} - k_{2}) - \eta (k_{2} - k_{3}) \}.
\nonumber
\end{eqnarray}
where Re[ $ x $ ] stands for the real part of $ x $.

   We consider the expectation value of ${\cal H}$ for the state
\begin{equation}
| \{ n_{k} \}  > = \prod_{k} (n_{k} ! )^{- \frac{1}{2}}
(a_{k}^{\dagger})^{n_{k}} | 0 >
\end{equation}
which is an eigenstate of ${\cal H}_{2}$. Then we get
\begin{eqnarray}
E &=& E_{0}+<{\cal H}_{2}>+<{\cal H}_{4}>, \\
<{\cal H}_{2}> &=& \sum_{k} <n_{k}> S \{ \eta (0) - {\rm Re} [\eta (k)] \},
\nonumber \\
<{\cal H}_{4}>  &=& - \frac{1}{2 N}
\Big(
\sum_{k} \frac{<n_{k}^{2}> - <n_{k}>}{2}
\Big\{ 2 \eta (0) -  \eta (k) - \eta ( - k )
\Big\}
\nonumber \\
& & +
\sum_{k,k^{\prime}(k \ne k^{\prime})}
<n_{k}n_{k^{\prime}}>
\Big\{ \eta (0) + \eta (k - k^{\prime}) - \eta (k) - \eta (k^{\prime})
\Big\}
\Big) ,
\nonumber
\end{eqnarray}
where $n_{k} = a_{k}^{\dagger} a_{k}$.
Magnetization in the $z$-direction is given
by $SN- \sum_{k} a_{k}^{\dagger} a_{k}$,
so the zero-magnetization condition is
\begin{equation}
SN- \sum_{k} a_{k}^{\dagger} a_{k} = 0.
\end{equation}

   We assume here that $n_{k} $ is the Bose distribution;
the entropy and the free energy are respectively written as follows:
\begin{equation}
({\rm entropy}) = \sum_{k} \{ (1+ n_{k}) \ln (1+ n_{k}) - n_{k} \ln n_{k} \},
\end{equation}
\begin{equation}
F = E  - T \times ({\rm entropy}).
\end{equation}
Moreover $<n^{2}_{k}>$ which expresses the expectation value of $n_{k}^{2}$
is $ 2 \tilde{n}_{k}^{2} + \tilde{n}_{k} $
in terms of $\tilde{n}_{k}$ ($ = <n_{k}>$).
We want to know $\tilde{n}_{k}$
which minimizes the free energy
under the constraint condition of zero magnetization, so we
introduce the Lagrange multiplier $\mu$ and
minimize the following quantity $W$:
\begin{equation}
W= F - \mu ( \sum_{k} \tilde{n}_{k} - S N ).
\end{equation}
{}From $\partial W / \partial \tilde{n}_{k} = 0 $
the Bose-Einstein distribution
\begin{equation}
\label{distribution}
\tilde{n}_{k} = \frac{1}{e^{ \beta ( \varepsilon (k) - \mu )}-1}
\end{equation}
is reproduced where
$\varepsilon (k) = \partial E  / \partial \tilde{n}_{k}$
and $\beta = T^{-1}$. From $\partial W / \partial \mu = 0 $
we have the self-consistent condition
\begin{equation}
\label{sc_equation_of_0_mag}
S = \frac{1}{N} \sum_{k} \tilde{n}_{k}
\end{equation}
which determines the chemical potential $\mu$.

 Using the rotational averaging, we obtain the static susceptibility;
\begin{equation}
\label{susceptibility}
\chi = \frac{\beta}{3 N} \sum_{k} (\tilde{n}_{k}^{2}+\tilde{n}_{k}).
\end{equation}

\eqreset
\section{Quadratic Theory}\label{quad}

  In this section we consider
the first two terms up to the quadratic term of operators
in the Holstein-Primakoff transformed Hamiltonian (\ref{hamiltonian2}).
Then the dispersion relation is written as follows:
\begin{equation}
\varepsilon (k) = J S \sum_{n=1}^{N - 1}
[\frac{\frac{\pi}{N}}{\sin (\frac{\pi n }{N})}]^{p} \{ 1 - \cos (kn) \} .
\end{equation}
Here we take the thermodynamic limit $N \rightarrow \infty$;
the dispersion relation is rewritten to
\begin{equation}
\varepsilon (k) = 2 J S \sum_{n=1}^{\infty}
 \frac{ 1 - \cos (kn) }{n^{p}} .
\end{equation}
The Bose-Einstein integral function gives us the dominant term
of the dispersion for any positive and small $k$
for arbitrary $p$ ($>1$).
\[
\varepsilon (k) \simeq \left \{
\begin{array}{rl}
  J S \zeta (p -2) k^{2} &\quad\mbox{$(p > 3)$} \\
- J S k^{2} \ln k  &\quad\mbox{$(p = 3)$} \\
J S \omega (p) k^{p -1} &\quad\mbox{$(1 < p < 3)$}
\end{array}\right.
\]
\begin{equation}
\omega (p) \equiv \frac{\pi}{\Gamma (p) \cos [\pi (p -2)/2]}.
\nonumber
\end{equation}

   In the region $p \geq 2$, we can determine the chemical potential $\mu$
from the self-consistent condition (\ref{sc_equation_of_0_mag}).
We have no Bose condensation
which breaks this condition (\ref{sc_equation_of_0_mag}).
The satisfaction of this condition means that
the system has no ordering transition.
We use the determined $\mu$
to calculate the susceptibility and the specific heat
at low temperature for $p \geq 2$.
The continuum approximation of the state density is valid
in the region $p \geq 2$;
then the self-consistent condition (\ref{sc_equation_of_0_mag})
is rewritten to
\begin{equation}
\label{SC_eq_in_DOP}
S = \frac{1}{\pi} \int_{0}^{\pi}
\frac{dk}{ e^{\beta \varepsilon ( k ) + v } - 1 }
= \frac{1}{\pi} \int_{0}^{\varepsilon (\pi)} \frac{dk}{d \varepsilon}
\frac{d \varepsilon }{e^{\beta  \varepsilon + v   } -1 }
\end{equation}
where $v=-\beta \mu$.

    First we consider the region $p > 3$.
{}From the dispersion relation and
the self-consistent condition (\ref{SC_eq_in_DOP}):
\begin{eqnarray}
S &=& \frac{1}{2 \pi \sqrt{J S \zeta (p -2)}}
\int_{0}^{\varepsilon (\pi)}
\frac{ \varepsilon^{-\frac{1}{2}}
d \varepsilon }{e^{\beta \varepsilon + v}-1} \nonumber \\
&\simeq& \frac{1}{2 \pi \sqrt{\beta J S \zeta (p -2)}}
\int_{0}^{\infty} \frac{x^{-\frac{1}{2}} dx}{e^{x+v}-1} \nonumber \\
&=& \frac{1}{2 \pi \sqrt{\beta J S \zeta (p -2)}}
\Gamma (\frac{1}{2}) \left( \Gamma (\frac{1}{2}) v^{-\frac{1}{2}}
+ \zeta (\frac{1}{2}) + \cdots  \right),
\end{eqnarray}
we obtain $v$ as follows:
\begin{equation}
v^{-1} =  4 \zeta (p -2) S^{3} \beta J
\end{equation}
where the Bose-Einstein integral function is used.
{}From eq. (\ref{susceptibility}) the susceptibility in this region
is calculated as follows:
\begin{equation}
\label{susceptibility_ESRR}
\chi \simeq \frac{2 S^{4} \zeta (p -2)}{3} \beta^{2} J
\end{equation}

   In the same way, we have
\[
v^{-1} \simeq \left \{
\begin{array}{rl}
\left(
S^{p} [\kappa (p)]^{p -1} \omega (p) \beta J
\right)^{\frac{1}{p -2}}
&\quad\mbox{$(2< p <3)$} \\
\exp ( \beta J S^{2} \pi^{2})
&\quad\mbox{$(p =2)$}
\end{array}\right.
\]
where $\kappa (p) \equiv (p -1) \sin [\pi / (p -1)]$;
then we have \[
\chi = \left \{
\begin{array}{rl}
\frac{p -2}{3 J (p -1)}
[ \omega (p) ]^{\frac{1}{p -2}}
\left\{ S^{2} \kappa (p) \beta J \right\}^{\frac{p -1}{p -2}}
&\quad\mbox{$(2< p <3)$} \\
\frac{1}{3 J S \pi^{2}}
\exp ( \beta J S^{2} \pi^{2})
&\quad\mbox{$(p =2)$}
\end{array}\right.
\]

   In this section the free energy per site is given as follows:
\begin{equation}
f \equiv \frac{F}{N}
= \frac{E_{0}}{N} + \frac{1}{N} \sum_{k} \tilde{n}_{k} \varepsilon (k)
- \frac{T}{N} \sum_{k} \{ (1+ \tilde{n}_{k}) \ln (1+ \tilde{n}_{k})
- \tilde{n}_{k} \ln \tilde{n}_{k} \}
\end{equation}
In the limit $N \rightarrow \infty$, we have
\begin{equation}
f = e_{0}
+ S \mu + \frac{T}{\pi} \int_{0}^{\varepsilon (\pi)} \frac{dk}{d\varepsilon}
\ln (1- e^{ - \beta \varepsilon - v }) d\varepsilon
\end{equation}
where $E_{0}/N \rightarrow e_{0}$ ( $= - J S \zeta (p) $\ )
as $N \rightarrow \infty$.
{}From the dispersion relation the free energy is obtained as follows:
\[
\frac{f-e_{0}}{T} \simeq \left \{
\begin{array}{rl}
-\frac{\zeta (3/2)}{\sqrt{2\pi \zeta(p -2)}} \sqrt{\frac{T}{2 S J}}
&\quad\mbox{$(p > 3 )$} \\
- \pi^{-1} \lambda (p) [S \omega (p)]^{\frac{1}{1-p}}
(\frac{T}{J})^{\frac{1}{p -1}} &\quad\mbox{$(1< p <3)$} \\
\end{array}\right.
\]
where $\lambda (p) \equiv \Gamma \Bigl( 1/ ( p -1 ) \Bigr)
\zeta \Bigl(p/ ( p -1 ) \Bigr) / (p -1)$.
Then we can calculate the specific heat per site; we have
\[
c \simeq \left \{
\begin{array}{rl}
\frac{3 \zeta (3/2)}{4 \sqrt{2\pi \zeta(p -2)}} \sqrt{\frac{T}{J}}
&\quad\mbox{$(p > 3 )$} \\
\frac{p}{\pi (p -1)^{2}} \lambda (p) [\omega (p)/2]^{\frac{1}{1-p}}
(\frac{T}{J})^{\frac{1}{p -1}}
&\quad\mbox{$(1< p <3)$} \\
\end{array}\right.
\]
for $S=1/2$.

         On the other hand, we cannot determine
the chemical potential $\mu$ in the region $1 < p < 2$
as can do in the region $p \geq 2 $.
This is because the zero-magnetization condition is broken
by the Bose condensation. Instead we can know
the critical temperature $T_{\rm c}$ (=$1/\beta_{\rm c}$)
of spontaneous magnetization
by the estimation of the critical temperature of the Bose condensation.
The critical temperature
in the region where $p$ is less than 2 and
where $p$ is in the vicinity of 2 can be obtained
from the constant terms of the Bose-Einstein integral function.
We should use $\varepsilon (k) = J S
\{ \omega (p) k^{p -1} + \zeta (p -2) k^{2} \}  $
as the dispersion relation
to obtain a better estimation of $T_{\rm c}$ near $p = 2$; then we have
\begin{equation}
\label{critical_T_2}
T_{\rm c} = J S \omega (p) [\tau (p)]^{p -1}_{,}
\end{equation}
\begin{displaymath}
\tau (p) =  \tau_{0} (p) +
\frac{(4 - p) \Gamma (\frac{4-p}{p -1})
\zeta (\frac{4-p}{p -1}) \zeta (p -2)}{(p -1) \omega (p)
\Gamma (\frac{1}{p - 1}) \zeta (\frac{1}{p - 1})}
\{ \tau_{0} (p) \}^{4-p}_{,}
\end{displaymath}
\begin{displaymath}
\tau_{0} (p)
= \frac{\pi S (p -1)}{\Gamma (\frac{1}{p -1}) \zeta (\frac{1}{p -1})}.
\end{displaymath}
We note that $\varepsilon (k) \rightarrow \varepsilon (\pi)$ for any $k >$  2
as $p \rightarrow 1+$.
The critical temperature for $p$ which is greater than 1
and which is in the vicinity of 1 can be estimated as follows;
\begin{equation}
\label{critical_T_1}
T_{\rm c} = \frac{(2^{p} - 1 ) \zeta(p) }{2^{p -1} \ln 3} J
\end{equation}
for $S = 1/2$.

\eqreset
\section{Quartic Theory}\label{quart}

    In this section we consider the first three terms up to the quartic term
of operators in the Hamiltonian (\ref{hamiltonian2}).
Because the dispersion relation is given by
$\varepsilon (k) = \partial E / \partial \tilde{n}_{k}$, we have
\begin{equation}
\varepsilon(k) = \sum_{n=1}^{N - 1} J
[\frac{\frac{\pi}{N}}{\sin (\frac{\pi n }{N})}]^{p} \{ 1 - \cos (kn) \}
[S - \frac{1}{N} \sum_{k^{\prime}} \tilde{n}_{k^{\prime}}
+ \frac{1}{N} \sum_{k^{\prime}} \tilde{n}_{k^{\prime}} \cos( k^{\prime} n )].
\end{equation}
The even function $\tilde{n}_{k}$ in respect of $k$ is expanded
into the following Fourier series:
\begin{eqnarray}
\tilde{n}_{k} &=& \frac{f(0)}{2} + \sum_{m=1}^{\infty} f(m) \cos(km), \\
f(m) &=& \int_{-\pi}^{\pi} \frac{dk}{\pi} \tilde{n}_{k} \cos(km), \
 (m=0,1,2,\cdots).
\label{fourier_coefficient}
\end{eqnarray}
The dispersion $\varepsilon(k)$ in the thermodynamic limit
$N \rightarrow \infty$ is expressed by
\begin{equation}
\label{dispersion_of_4th_order}
\varepsilon(k) = J \sum_{n=1}^{\infty} \frac{1-\cos(kn)}{n^{p}} f(n)
\end{equation}
where we use the self-consistent condition of zero magnetization
(\ref{sc_equation_of_0_mag}) in this limit i.e.  $S=f(0)/2$.

    So the problem of calculating physical quantities is reduced to
obtaining the dispersion and the distribution
which satisfy the three eqs. (\ref{distribution}),
(\ref{fourier_coefficient}) and (\ref{dispersion_of_4th_order})
under the self-consistent condition of zero magnetization.
For arbitrary $p$ ($\ge 2$)
we can obtain $\varepsilon (k)$ and $v$
by the following equations:
\begin{equation}
\label{iteration_1}
\varepsilon (k) = \frac{2 J}{\pi} \sum_{n=1}^{\infty}
\int_{0}^{\pi} dq \frac{\left( 1 - \cos (k n)\right) \cos (q n)}{n^{p}
(e^{\beta \varepsilon (q) +v} - 1) }
\end{equation}
\begin{equation}
\label{iteration_2}
S \pi = \int_{0}^{\pi} \frac{dk}{e^{\beta \varepsilon (k) +v} - 1}
\end{equation}
The dispersion and chemical potential
which are obtained from the iteration of eqs. (\ref{iteration_1})
and (\ref{iteration_2})
give us physical quantities.

    Fortunately we can make an analytical treatment for $p=2$;
it is shown in the rest of this section.
Equation (\ref{dispersion_of_4th_order}) for $p=2$ is differentiated twice;
we have
\begin{equation}
\label{dif_eq}
\frac{d^{2}\varepsilon}{dk^{2}} = J \tilde{n}_{k} - J S.
\end{equation}
The introduction of a function $g(k)=d \varepsilon / d k$ and
the integration of the differentiating equation (\ref{dif_eq})
give us the following equation:
\begin{equation}
\label{solution_of_dif_eq}
g^{2} = \frac{2 J}{\beta} \ln
\left| \frac{1-e^{-\beta \varepsilon - v}}{1-e^{- v}} \right|
-2 J S \varepsilon
\end{equation}
where the initial conditions $g = \varepsilon = 0 $ at $ k = 0 $ are used.
{}From eq. (\ref{solution_of_dif_eq}), we use the equation:
\begin{equation}
\label{changing_variable}
dk =
\frac{ d\varepsilon}{\sqrt{\frac{2 J}{\beta} \ln
|\frac{1-e^{-\beta \varepsilon - v}}{1-e^{- v}}|
-2 J S \varepsilon}   }
\end{equation}
to change an integral variable from $k$ to $\varepsilon$
in the integrations.

   The substitution of $ \pi $ for $k$ in eqs.
(\ref{dispersion_of_4th_order}) and (\ref{solution_of_dif_eq}) gives
\begin{equation}
\varepsilon (\pi) = \frac{J S \pi^{2}}{2} + O (v \ln \frac{1}{v}),
\end{equation}
\begin{equation}
e^{v} = 1 -
\frac{1 - e^{- \beta \varepsilon (\pi)}}{1-e^{S \beta \varepsilon (\pi) }}
\end{equation}
respectively. At low temperature, then we have
\begin{equation}
v \simeq \exp ( - \frac{ \beta J S^{2} \pi^{2} }{2} ).
\end{equation}

   From eqs. (\ref{changing_variable}) and (\ref{susceptibility})
the susceptibility at low temperature of the HS model is calculated as
$\chi \simeq (\beta/6) \sqrt{2/ \beta J \pi}
\exp (\beta J S^{2} \pi^{2}/2 )$.
For  $S = 1/2$ the susceptibility is obtained as follows:
\begin{equation}
\label{susceptibility_quartic_theory}
\chi \simeq \frac{\beta}{6} \sqrt{\frac{2}{\beta J \pi}}
\exp (\frac{\beta J \pi^{2}}{8} ).
\end{equation}

    The free energy per site in this section is given as follows:
\begin{eqnarray}
f &=& \frac{1}{4} \sum_{n=1}^{\infty} \frac{J}{n^{p}} \{ (f(n) \}^{2}
- J S \sum_{n=1}^{\infty} \frac{f(n)}{n^{p}}
+ \mu S \nonumber \\
& & + \frac{1}{\beta \pi} \int_{0}^{\varepsilon (\pi)}
\frac{dk}{d\varepsilon}
\ln (1- e^{- \beta \varepsilon - v }) d\varepsilon.
\end{eqnarray}
Then we obtain the following dominant term of the specific heat per site:
\begin{equation}
\label{specific_heat_quartic_theory}
c   \simeq \frac{2}{3} (\frac{T}{J})
\end{equation}
for $p = 2$ and $S=1/2$.

\eqreset
\section{Discussion}\label{discus}

  In this paper we have investigated the one-dimensional quantum
Heisenberg model with long-range ferromagnetic interactions
by the modified spin-wave approximation.
This approximation makes it possible to treat the cases
not only of the special values of $p$.

     We can apply the modified spin-wave theory to the NN Heisenberg model;
its validity has already been checked.\cite{mspin}
The limit $p \rightarrow \infty$ shifts
both the quadratic and quartic theory in this paper
straightforward to the cases of the NN Heisenberg model.

     The susceptibility and the entropy density of
the ferromagnetic HS model
are obtained by Haldane\cite{Haldane2} as follows:
\begin{equation}
\label{susceptibility_HS_model}
\chi
= \frac{\beta}{2 \pi} \int_{0}^{\pi/2} d v \exp ( - 2 \beta J h )
\simeq \frac{\beta}{4} \sqrt{\frac{2}{\beta J \pi}}
\exp (\frac{\beta J \pi^{2}}{8} ),
\end{equation}
\begin{equation}
\label{entropy_density_HS_model}
s
= \frac{2}{\pi} \int_{0}^{\pi/2} d v \{ \ln [2 \cosh (\beta J h)] - \beta J h
\tanh (\beta J h) \}
\end{equation}
respectively, where $h (v) = [v^{2} - (\pi/2)^{2}]/4$.
{}From eq. (\ref{entropy_density_HS_model})
the specific heat per site is obtained;
we have
\begin{equation}
\label{specific_heat_HS_model}
c  = \frac{2}{3} (\frac{T}{J}) + \cdots.
\end{equation}

    Both the results of the specific heat in the quadratic theory and
that in the quartic theory are the same as this specific heat
(\ref{specific_heat_HS_model}).
Our result of the susceptibility from the quadratic theory
is not the same as the expression (\ref{susceptibility_HS_model})
but is in agreement with it from the point of view
that both have exponential divergence at low temperature.
So it is reasonable that the quadratic theory is qualitatively valid.

    From our results in the quadratic theory
we can divide the region of $p$
which determines the strength of long-range interactions into two parts.
One is the region where
temperature-dependence except for coefficients of physical quantities is
the same as the case of the NN Heisenberg model at low temperature.
The other is the region where at low temperature
it is different from the case of the NN Heisenberg model.
We will call the former region
the effectively short-range region (ESRR)
and we will call the latter region
the essentially long-range region (ELRR).
In the region $p > 3$
the susceptibility  has the same power of $T$
as that of the NN Heisenberg model
in spite that the model in this region has long-range interactions.
The specific heat also has the same power of $T$.
In the region $2<p<3$
the susceptibility and the specific heat have behaviors of power of $T$
as temperature goes to zero. The powers of temperature in the expressions
of the susceptibility and of the specific heat, however, are changed
as $p$ goes from 3 to 2.
Especially the power of the susceptibility goes to infinity
as $p \rightarrow 2+$;
the susceptibility for $p=2$ has exponential divergence
at low temperature.
Then the region $p > 3$ is the ESRR and the region $p < 3$ is the ELRR.
The authors believe that the model for $p=3$ is in the ELRR.
This is because there is a possibility that
the susceptibility has the divergence of $1/T^{2}$
with a factor of correction $\ln (1/T)$.
The conclusion of the quadratic theory is summarized
in Fig. 1.

    Until now no approaches to the estimation
of the critical temperature in the region $1 < p < 2$
have been known.
So we cannot compare our results with others.
Our estimation of $T_{\rm c}$, however,
is convincing from the following two points of view.
One is that $T_{\rm c}$ of the expression (\ref{critical_T_2}) vanishes
as $p \rightarrow 2$.
The other is that
$T_{\rm c} $ of the expression (\ref{critical_T_1})
is divergent as $p \rightarrow 1$.
This has no contradiction to the fact that
it takes an infinite amount of energy to make one-magnon state from the vacuum.

    Our result (\ref{susceptibility_quartic_theory}) of  the susceptibility
in the quartic theory are in good agreement
with the expression (\ref{susceptibility_HS_model});
the only difference is the constant factor $2/3$.
This constant factor comes from the point that
the modified spin-wave theory is  not rotationally invariant.
This constant factor reminds us of the difference between
the susceptibility from the modified spin-wave theory and
that from the Schwinger boson mean field theory.\cite{AA}
Results from the Schwinger boson mean field theory will be published elsewhere.
There it will be shown that almost all the same discussion
in this paper is made again.
The only difference will be the improvement of this constant factor
in the susceptibility.

\vspace{3mm}
\begin{flushleft}
{\bf Acknowledgement}
\end{flushleft}
\vspace{2mm}

  One of the authors (M.T.) thanks to Duncan Haldane
for stimulating discussion at Aspen Center for Physics in 1991.

\renewcommand{\theequation}{\Alph{section}.\arabic{equation}}

\appendix{}

  The Bose-Einstein integral function
\begin{equation}
F(\alpha,v)= \frac{1}{\Gamma(\alpha)}
\int_{0}^{\infty} \frac{x^{\alpha -1}dx}{e^{x+v}-1}
= \frac{e^{- v}}{1^{\alpha}} + \frac{e^{-2 v}}{2^{\alpha}}  + \cdots ,
\end{equation}
is often used in this paper. Analytical property of this function near $v=0$
is known; \ we have
\begin{displaymath}
F(\alpha,v)= \Gamma(1-\alpha) v^{\alpha -1}
+ \sum_{n=0}^{\infty} \frac{\zeta(\alpha-n)}{n!} (- v)^{n} \ \ \
(\alpha \notin \mbox{\boldmath $N$}),
\end{displaymath}
\begin{displaymath}
F(\alpha,v)= \frac{(- v)^{\alpha -1}}{(\alpha -1)!}
\{ \sum_{r=1}^{\alpha -1} \frac{1}{r} - \ln v \}
+ \sum_{n \ne \alpha -1} \frac{\zeta(\alpha-n)}{n!} (- v)^{n} \ \ \
(\alpha \in \mbox{\boldmath $N$}),
\end{displaymath}
where $\zeta (\alpha)$ is Riemann's zeta function.
We need the summation
\begin{equation}
\label{sum_of_dispersion}
\sum_{n=1}^{\infty} \frac{1-\cos(kn)}{n^{\alpha}},
\end{equation}
to obtain the dispersion relation.
We can calculate the dominant term of this summation for
small $k$ using $F(\alpha,v)$; \
(\ref{sum_of_dispersion}) is expressed by \
$\zeta (\alpha) - {\rm Re} [ F(\alpha , - i k) ]$. \
The following integrations are very useful to obtain the susceptibility
and the free energy:
\begin{equation}
\int_{0}^{\infty} \frac{x^{\alpha -1} e^{x+v} dx}{(e^{x+v}-1)^{2}}
= \Gamma (\alpha) F(\alpha -1,v),
\end{equation}
\begin{equation}
\int_{0}^{\infty}
x^{\alpha -1} \ln (1 - e^{- x - v}) d x
= - \Gamma (\alpha) F(\alpha + 1,v)
\end{equation}
respectively.

\newpage
{\large \bf Figure Captions}\\

{\bf Fig.1 } \hspace{5mm} The critical temperature
in the region $ 1 < p < 2  $ and
the exponent of the susceptibility in the region $ p \ge 2 $.
The solid line stands for the exponent.
The exponent for $p = 3$ isn't obtained in this paper.
The critical temperature is expressed by the broken line.
Numerical results link the analytical results (\ref{critical_T_2})
and (\ref{critical_T_1}).


\begin{references}
  \bibitem{Haldane1}
      F. D. M. Haldane:
         Phys.Rev.Lett. {\bf   60}(1988)635.
  \bibitem{Shastry}
      B. S. Shastry:
         Phys.Rev.Lett. {\bf   60}(1988)639.
  \bibitem{Haldane2}
      F. D. M. Haldane:
         Phys.Rev.Lett. {\bf   66}(1991)1529.
  \bibitem{Ito}
      K. R. Ito:
         J.Stat.Phys. {\bf   29}(1982)747.
  \bibitem{Dyson}
      F. J. Dyson:
         Commun.Math.Phys. {\bf   12}(1969)91.
  \bibitem{Simanek}
      E. \v{S}im\'{a}nek:
         Phys.Lett. {\bf   A 119}(1987)477.
  \bibitem{Romano}
      S. Romano:
         Phys.Rev. {\bf   B 40}(1989)4970.
  \bibitem{Fisher}
      M. E. Fisher, S. Ma and B. G. Nickel:
         Phys.Rev.Lett. {\bf   29}(1972)917.
  \bibitem{MSuzuki}
      M. Suzuki:
         Phys.Lett. {\bf   42 A}(1972)5.
  \bibitem{mspin}
      M. Takahashi:
         Prog.Theor.Phys.Suppl.{\bf 87}(1986)233.
  \bibitem{TY1}
      M. Takahashi and M. Yamada:
      J.Phys.Soc.Jpn.  {\bf 54}(1985)2808.
  \bibitem{TY2}
      M. Yamada and M. Takahashi:
      J.Phys.Soc.Jpn.  {\bf 55}(1986)2024.
  \bibitem{AA}
      D. P. Arovas and A. Auerbach:
         Phys.Rev. {\bf  B 38 }(1988)316.
\end{references}
\end{document}